\documentclass[aps,pra,preprint,showpacs,floatfix,a4paper,superscriptaddress]{revtex4-1}
\usepackage{amsmath}
\usepackage[dvips]{graphicx}
\usepackage{epsfig}
\usepackage{tabularx}
\usepackage{color}
\usepackage[pdfpagemode=UseNone,colorlinks=true,linkcolor=blue,citecolor=blue]{hyperref}

\renewcommand{\lor}{\mathcal{L}}
\newcommand{\wm}{\omega_{\mathrm{m}}}
\newcommand{\gm}{\gamma_{\mathrm{m}}}
\newcommand{\gef}{\gamma_{\mathrm{eff}}}

\newcommand{\wef}{\omega_{\mathrm{eff}}}
\newcommand{\woef}{\omega^0_{\mathrm{eff}}}
\newcommand{\we}{\omega_{\mathrm{e}}}

\begin{document}

\title{Detection of weak stochastic force in a parametrically stabilized micro opto-mechanical system}

\author{A. Pontin}
\affiliation{Istituto Nazionale di Fisica Nucleare (INFN), Gruppo Collegato di Trento, I-38123 Povo, Trento, Italy}
\affiliation{Dipartimento di Fisica, Universit\`a di Trento, I-38123 Povo, Trento, Italy}
\author{M. Bonaldi}
\affiliation{Institute of Materials for Electronics and Magnetism, Nanoscience-Trento-FBK Division,
 38123 Povo, Trento, Italy}
\affiliation{INFN, Gruppo Collegato di Trento, Sezione di Padova, 38123 Povo, Trento, Italy}
\author{A. Borrielli}
\affiliation{Institute of Materials for Electronics and Magnetism, Nanoscience-Trento-FBK Division,
 38123 Povo, Trento, Italy}
\affiliation{INFN, Gruppo Collegato di Trento, Sezione di Padova, 38123 Povo, Trento, Italy}

\author{F. S. Cataliotti}
\affiliation{Dipartimento di Fisica e Astronomia, Universit\`a di Firenze, Via Sansone 1, I-50019 Sesto Fiorentino (FI), Italy}
\affiliation{European Laboratory for Non-Linear Spectroscopy (LENS), Via Carrara 1, I-50019 Sesto Fiorentino (FI), Italy}
\affiliation{INFN, Sezione di Firenze}

\author{F. Marino}
\affiliation{INFN, Sezione di Firenze}
\affiliation{CNR-INO, L.go Enrico Fermi 6, I-50125 Firenze, Italy}

\author{G. A. Prodi}
\affiliation{Istituto Nazionale di Fisica Nucleare (INFN), Gruppo Collegato di Trento, I-38123 Povo, Trento, Italy}
\affiliation{Dipartimento di Fisica, Universit\`a di Trento, I-38123 Povo, Trento, Italy}

\author{E. Serra}
\affiliation{Istituto Nazionale di Fisica Nucleare (INFN), Gruppo Collegato di Trento, I-38123 Povo, Trento, Italy}
\affiliation{Dept. of Microelectronics and Computer Engineering /ECTM/DIMES, Delft University of Technology, Feldmanweg 17, 2628 CT  Delft, The Netherlands}

\author{F. Marin}
\email[Electronic mail: ]{marin@fi.infn.it}
\affiliation{Dipartimento di Fisica e Astronomia, Universit\`a di Firenze, Via Sansone 1, I-50019 Sesto Fiorentino (FI), Italy}
\affiliation{European Laboratory for Non-Linear Spectroscopy (LENS), Via Carrara 1, I-50019 Sesto Fiorentino (FI), Italy}
\affiliation{INFN, Sezione di Firenze}

\date{\today}
\begin{abstract}
Measuring a weak force is an important task for micro-mechanical systems, both when using devices as sensitive detectors and, particularly, in experiments of quantum mechanics. The optimal strategy for resolving a weak stochastic signal force on a huge background (typically given by thermal noise) is a crucial and debated topic, and the stability of the mechanical resonance is a further, related critical issue. We introduce and analyze the parametric control of the optical spring, that allows to stabilize the resonance and provides a phase reference for the oscillator motion, yet conserving a free evolution in one quadrature of the phase space. We also study quantitatively the characteristics of our micro opto-mechanical system as detector of stochastic force for short measurement times (for quick, high resolution monitoring) as well as for the longer term observations that optimize the sensitivity. We compare a simple, na$\ddot{\textrm{i}}$ve strategy based on the evaluation of the variance of the displacement (that is a widely used technique) with an optimal Wiener-Kolmogorov data analysis. We show that, thanks to the parametric stabilization of the effective susceptibility, we can more efficiently implement Wiener filtering, and we investigate how this strategy improves the performance of our system. We finally demonstrate the possibility to resolve stochastic force variations well below 1$\%$ of the thermal noise.
\end{abstract}


\maketitle

\section{Introduction}

Micro- and nano-opto mechanical systems are the heart of refined force-sensing devices \cite{Gavartin2012,Miao2012,Poggio2013,Purdy2013}. Such systems exploit the huge susceptibility around the resonance of oscillators with excellent mechanical quality factor $Q$, combined with high sensitivity interferometric measurements. The latter are particularly efficient when the oscillator is embedded in an optical resonator with high optical quality factor, whose optical path depends on the oscillator coordinate. This kind of devices is useful both for practical applications, and in quantum optics experiments. In both cases, a frequent crucial task is detecting a weak variation of the external force (that we call signal force) on a strong background. For instance, in a quantum experiment, the signal can be due to quantum fluctuations in the radiation pressure, that are usually overwhelmed by background thermal noise (a significant exception is reported in Ref. \cite{Purdy2013}, that presents the first observation of the effect of radiation pressure shot noise on a macroscopic object). 

Due to the narrow width of the resonance and, consequently, of the useful sensitive band with respect to typical input force, it is meaningful to discuss the general problem of detecting a weak signal force with flat spectral density (white spectrum) in the presence of a white background force, taking into account a given sensitivity to the oscillator displacement (i.e., a flat readout noise spectrum). This can be performed na$\ddot{\textrm{i}}$vely by measuring the area of the resonance peak emerging from the displacement noise spectrum (or, equivalently, measuring the variance of the oscillator position after band-pass filtering around the resonance). With this estimator, the rate of improvement of the statistical uncertainty for increasing measurement time $t_{meas}$ depends on the correlation time $\tau_c$ of the oscillator motion, with a relative uncertainty scaling as $\sim\sqrt{\tau_c/t_{meas}}$. It seems therefore useful to decrease $\tau_c$, i.e., enhance the damping of the oscillator. However, the fluctuation-dissipation theorem implies that such operation would increase the spectral density of thermal noise. Improved results can instead be achieved by means of a cold damping, e.g. the optical cooling \cite{Kleckner06,Gigan06,Arcizet06}, that modifies the effective susceptibility and decreases the correlation time without introducing additional noise sources. This technique does not increase the signal-to-noise ratio of input excitations, because it changes the response to both signal and background force in the same way. However, as long as the cold damped peak still emerges from the displacement spectral noise, it allows a faster accumulation of statistically independent data bringing therefore, in a given measurement time, to a smaller final uncertainty in the variance of the oscillator motion.

An important remark is that the correlation time of the signal force is by hypothesis very short, therefore the statistics can in principle be much faster than what allowed by the oscillator motion. In other words, the variance of the displacement is not a very efficient indicator, and more refined data analysis can be profitable. In the case of stationary, white input the optimal approach to the measurement is provided by the Wiener-Kolmogorov filtering theory \cite{Kolmogorov,Wiener}. This technique requires the preliminary knowledge of the exact response function to the input force, and of the signal-to-noise ratio. While the second requirement can be relaxed with a sub-optimal but robust filter using a conservative estimate of the sensitivity \cite{Astone1990}, the accurate knowledge of the susceptibility is a crucial request. Such knowledge is not trivial for micro opto-mechanical systems, where the stability of the resonance is affected by several detrimental effects, e.g., thermal phenomena and relaxations of the mechanical oscillator, and above all by the same interaction with the radiation, both due to photothermal effect and to the opto-mechanical coupling. These considerations suggest that the direct measurement of the spectral peak area could be the only applicable strategy in several kinds of opto-mechanical systems, and techniques that reduce the effective coherence time of the oscillator  motion, such as cold damping or feedback, represent therefore a way to effectively improve the measurement capabilities of the system \cite{Gavartin2012}. However, it has been remarked that optimal resolution is not really improved in this way \cite{Tamayo2005,Vinante2013}, and that appropriate data filtering can completely replace these hardware techniques even in the case of non-stationary, non-Gaussian input \cite{Harris2013}. In spite of these correct remarks, the problem of the instability in the oscillator parameters and dynamics remains practically difficult to face, and the implementation of optimal analysis requires sophisticated technique of adaptive filtering. The experimental demonstration in Ref. \cite{Harris2013} keeps indeed short ($\sim$ms) measurement times. Therefore, even when willing to apply an efficient data analysis, as well as in several kinds of refined opto-mechanics experiments, stabilization and feedback techniques acting on the opto-mechanical system are crucial, and indeed this issue has been recently considered by few groups\cite{Antonio2012,Gavartin2013}.

In this work we present a micro opto-mechanical system that includes a parametric stabilization of the resonance by controlling the optical spring. We have proposed and demonstrated this technique in a recent work \cite{Pontin2013}, where the control allows to prevent instability in a parametrically modulated opto-mechanical system, thus yielding strong mechanical squeezing. Here we study the characteristics of our system as detector of stochastic force for short measurement times (for quick, high resolution monitoring) as well as for long $t_{meas}$, thus optimizing the sensitivity. We show that, thanks to the stabilization of the effective susceptibility, we can more efficiently implement Wiener filtering and investigate how this strategy improves the performance of our system.

The article is organized as follows. In Section II we describe the theoretical models for the opto-mechanical interaction, the parametric control of the oscillator, and the strategies exploitable to measure the stochastic force acting on the oscillator; in Section II we describe our experimental setup and the measurements; after the Conclusions, in the Appendix we derive the theoretical expressions for the relative uncertainty and discuss the effect of a cutoff in the measured spectra.

\section{Model}

\subsection{Opto-mechanical interaction}

In this section we recall some basic features of the opto-mechanical interaction. We consider an optical cavity where the resonance frequency depends on an effective coordinate $x$, that is kept at its rest position $x=0$ by elastic forces. The system can be sketched as a linear cavity with a rigid oscillating mirror (Fig. \ref{schema}) having position $x$, mass $M$, resonance angular frequency $\wm$, damping rate $\gm$ and susceptibility $\chi = 1/M(\wm^2-\omega^2+\mathrm{i} \omega \gm)$. The radiation pressure provides a force acting on the mirror, that depends on the detuning $\Delta= \omega_L - \omega_c$ between the input radiation at frequency $\omega_L$ and the cavity resonance at $\omega_c$. Since the latter depends on $x$, radiation pressure gives a position-dependent force that can be accounted for by defining an effective susceptibility. Its expression is given by\cite{Arcizet06,Genes2008}
\begin{equation}
\chi _{\mathrm{ eff}}(\omega )^{-1}=M\left[\omega_\mathrm{m}^{2}-\omega^{2}+\mathrm{i}\omega \gamma _\mathrm{m}+\frac{|G|^2\,\Delta\,\omega _\mathrm{m}}{\bigl(\kappa +\mathrm{i}\omega \bigr)^{2}+\Delta^{2}}\right] 
\label{chieff}
\end{equation}
where $\kappa$ is cavity decay rate and $|G|^2$ is the opto-mechanical coupling, proportional to the intracavity power.

The real part of $\chi_{\mathrm{eff}}$ can be viewed as a combined effect of the mechanical stiffness and an additional spring (\emph{optical spring})\cite{Braginsky1997}. The delay in the intracavity field build-up, originating a contribution to the imaginary part of $\chi_{\mathrm{eff}}$, causes a change in the oscillator damping that allows the optical cooling of its motion \cite{Kleckner06,Gigan06,Arcizet06}. 

For the case of our interest (\emph{bad cavity limit} $\kappa\gg \wm$, small detuning $\Delta\ll
  \kappa$, and $\omega\approx \wm$) the expression of optical spring constant simplifies to
\begin{equation}
K_{\mathrm{opt}} \approx -\frac{M |G|^2 \wm}{\kappa^2}\,\Delta
\label{Kopt}
\end{equation}
and the optical damping rate to
\begin{equation}
\gamma_{\mathrm{opt}} \approx \frac{2 K_{\mathrm{opt}}}{M \kappa}
\label{gammaopt}
\end{equation}
allowing to write the effective susceptibility as $\,\chi_{\mathrm{eff}}^{-1}=M\left(\omega_{\mathrm{eff}}^2-\omega^2+\mathrm{i}\omega\,\gamma_{\mathrm{eff}}\right)\,$ with $\,\,\gamma_{\mathrm{eff}}=\gm+\gamma_{\mathrm{opt}}\,\,$ and
\begin{equation}
\omega_{\mathrm{eff}} = \sqrt{\wm^2-K_{\mathrm{opt}}/M}\simeq \wm + \frac{|G|^2}{2\kappa^2}\,\Delta \,.
\label{omegaeff}
\end{equation}
To our purpose, it is useful to underline that the frequency shift is approximately proportional to the detuning, and therefore a laser beam can be used to control it. Moreover, by varying the working point (detuning) we can choose the effective resonance width $\gef$ and stabilize it. On the other hand, we remark that in general the optical spring increases the uncertainty and instability of the opto-mechanical resonance frequency $\wef$ since it is influenced by the noise in the laser intensity (through $G$), in the laser frequency and in the cavity length (through $\Delta$). In addition, thermal effects due to the absorbed laser power can worsen the intrinsic stability of $\wm$.

\subsection{An oscillator with parametric control}

\begin{figure} [h]
\centering
\includegraphics[width=0.9\textwidth]{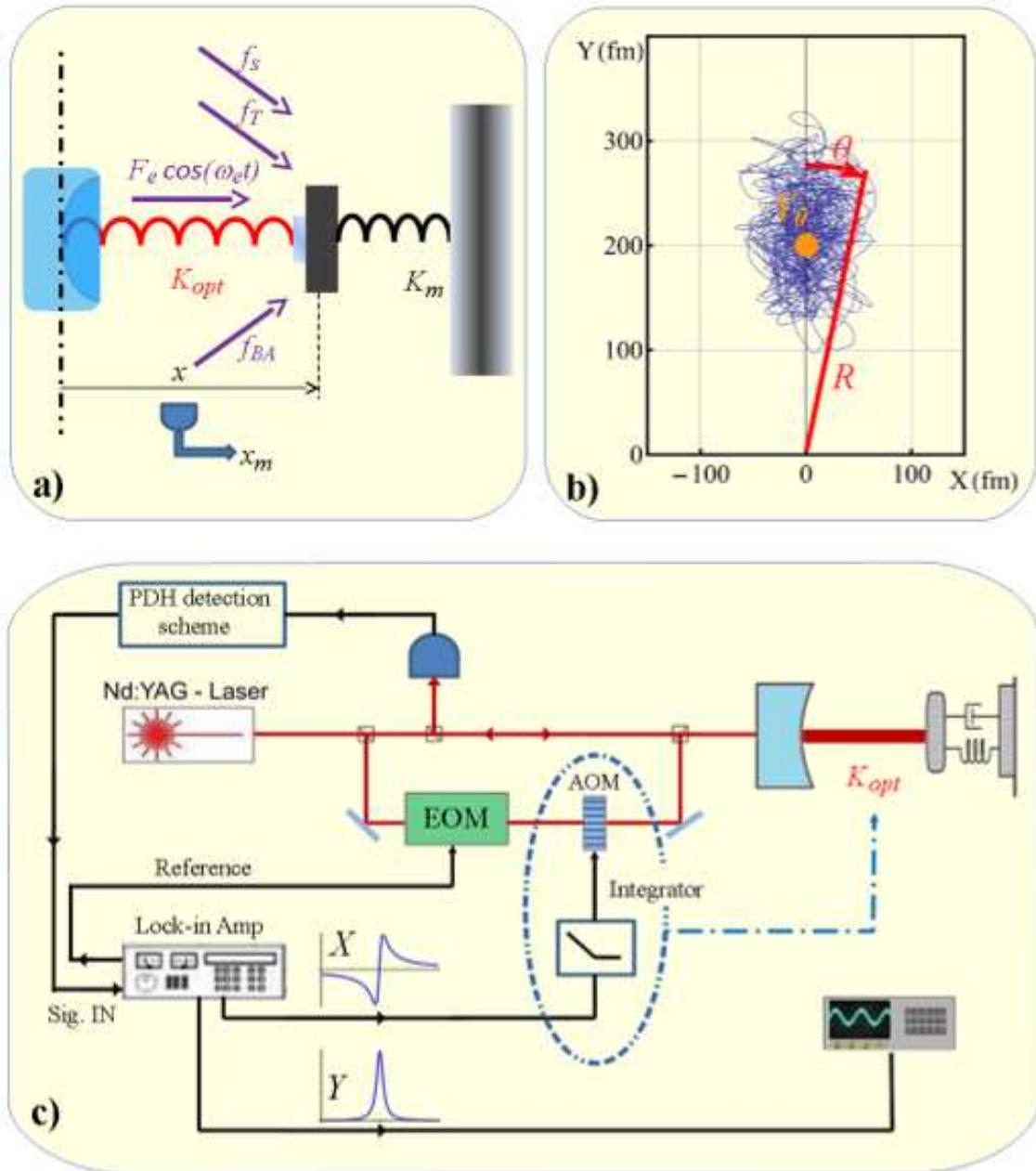}
\caption{(Color online) a) Conceptual scheme of the opto-mechanical system including measurement and force terms. b) Experimental measurement of the temporal evolution of the mechanical oscillator in the phase plane, in the configuration with active parametric control. c) Experimental setup. EOM: electro-optic intensity modulator; dash-dotted lines highlight the parametric control.}
\label{schema}
\end{figure}

A conceptual scheme of the experiment is shown in Fig. \ref{schema}a. We consider an opto-mechanical oscillator excited by stochastic signal force $f_s(t)$ and thermal noise force $f_T(t)$ at temperature $T$, with respective spectral densities $S_s$ and $S_T = 2k_B T M \gm$, as well as by a coherent oscillating force of constant amplitude $F_e \cos \we t$.  The oscillator position $x(t)$ is measured interferometrically by a first laser beam (\emph{signal beam}). The measurement noise $n(t)$ and the back-action force $f_{BA}(t)$ are considered uncorrelated, with white spectra $S_n$ and $S_{BA}$ bounded by
\begin{equation}
S_n\,S_{BA} \geq \hbar^2/4 \, .
\end{equation}
The evolution of the position $x(t)$ is governed by the stochastic equation
\begin{equation}
\ddot{x} + \gef \dot{x} + \wef^2 x = \frac{1}{M}\left[f_T + f_s + f_{BA} + F_e \cos \we t\right]
\label{eqx}
\end{equation}
and the result of the position measurement is $x_m(t) = x(t) + n(t)$.

The motion of the oscillator can be decomposed into two quadratures $X(t)$ and $Y(t)$ in a frame rotating at angular frequency $\we$, according to
\begin{equation}
x(t) = X(t) \cos \we t + Y(t) \sin \we t \, .
\end{equation}
Assuming $|\we-\wef|\ll \wef$, and $\gef \ll \wef$, the evolution equations for the two slowly-varying quadratures, derived from Eq. (\ref{eqx}), can be written as
\begin{subequations}
\label{eqXY} 
\begin{align}
\dot{X}+\frac{\gef}{2} X - \left(\wef-\we \right)Y = \frac{1}{M \we}\left[f^{(1)}_T + f^{(1)}_s + f^{(1)}_{BA} \right]    \\
\dot{Y}+\frac{\gef}{2} Y + \left(\wef-\we \right)X = \frac{1}{M \we}\left[f^{(2)}_T + f^{(2)}_s + f^{(2)}_{BA} + \frac{F_e}{2} \right] 
\end{align}
\end{subequations}
where the stochastic force terms have correlation functions $\langle f_a^{(i)}(t) f_a^{(j)} (t')\rangle = \delta_{ij}\delta(t-t') S_a/2 $ (i,j=1,2 and $"a" = "T", "s", "BA"$). In the experiment, the two quadratures are measured by sending $x_m(t)$ to a lock-in amplifier whose reference signal is derived from the oscillator modulating the coherent force $F_e$. The outputs of the lock-in are $X_m = X+n^{(1)}$ and $Y_m = Y + n^{(2)}$ with $\langle n^{(i)}(t) n^{(j)} (t')\rangle = \delta_{ij}\delta(t-t') S_n/2$.

The steady state solutions of Eqs. (\ref{eqXY}) are the usual components of the oscillator response, as a function of the frequency difference between resonance and excitation $\delta \omega = \wef-\we $:
\begin{subequations}
\begin{align}
\overline{X}\left( \delta \omega  \right) = \frac{F_e}{2}\frac{ \delta \omega  }{\frac{\gef^2}{4}+\left( \delta \omega  \right)^2}    \\
\overline{Y}\left( \delta \omega  \right) = \frac{F_e}{2}\frac{ \gef/2  }{\frac{\gef^2}{4}+\left( \delta \omega  \right)^2}  \, .
\end{align}
\end{subequations}
We remark that $\overline{X}$ is an odd function of $\delta \omega$, therefore it can be efficiently exploited to control and lock $\wef$. The $X_m$ quadrature is indeed integrated and sent to control the resonance frequency $\wef$ by modifying the optical spring constant (\emph{parametric control}). This is obtained in the experiment by acting on the detuning of a second laser beam (\emph{control beam}) according to
\begin{equation}
\omega_L(t) = \omega_L^0 - \int_{-\infty}^{t} \,\mathcal{G}(t,t')\, X_m(t') \mathrm{d}t'
\label{loop1}
\end{equation}
where $\omega_L^0$ is the initial detuning and the kernel $\mathcal{G}(t,t')$ is constant in the case of an integral feedback loop. 
Given that $\omega_L$ determines the effective frequency $\wef$ via Eq. (\ref{omegaeff}), we can write 
\begin{equation}
\wef(t) = \woef(t) - \int_{-\infty}^{t} \,\mathcal{\bar{G}}(t,t')\, X_m(t') \mathrm{d}t'
\label{loop}
\end{equation}
where $\woef(t)$ is the free-running opto-mechanical frequency and $\mathcal{\bar{G}} \propto \mathcal{G}$.
Eq. (\ref{gammaopt}) shows that, in the \emph{bad cavity} limit, the shift in the resonance frequency $\wef$ due to the opto-mechanical interaction is larger than the variation in the damping rate $\gef$, thus the latter can be neglected when considering small variations of $\Delta$ around the working point. We also remark that the control of the optical spring can be considered as a classical effect, and its noise neglected in a first-order treatment. In any case, such noise (for us, the radiation pressure noise of the control beam) can be included in $f_s$. 

At the purpose of analyzing the effect of the control loop, we first consider slow fluctuations in the opto-mechanical resonance frequency $\wef$, that can be treated as adiabatic changes of the system, keeping the validity of Eqs. (\ref{eqXY}). In Eq. (\ref{loop}) we replace $X = \overline{X}(\delta\omega) + \delta X$ and, considering small closed-loop fluctuations, we further take $\overline{X}(\delta \omega) \propto \delta \omega$. In the absence of drift in $\woef(t)$, the steady-state solution is $\delta \omega = 0$, i.e., $\wef = \we$ (long term drifts in $\woef(t)$ can be corrected by additional integrators, as in standard servo-loop systems). In the phase plane of a reference frame rotating at $\we$, the oscillator motion 
is now represented by a vector $\mathbf{R} = (X, Y)$ fluctuating around the average value $(0, Y_0)$ with $Y_0= \overline{Y}(0)=F_e/\gef$ (in Fig. \ref{schema}b we report an experimental example). The feedback loop corrects the fluctuations by counter-rotating $\mathbf{R}$ towards the $Y$ axis. If $\mathbf{R}$  remains close to  $(0, Y_0)$, i.e., if $\langle X^2 + (Y-Y_0)^2 \rangle \ll Y_0^2$, we can approximate the angle $\theta$ between $\mathbf{R}$ and the $Y$ axis with $\theta \approx X/Y_0$. 
In this limit, the feedback loop (that acts on $\theta$) just influence the fluctuations in the $X$ quadrature, leaving free $Y$ fluctuations. This is expressed by a linear expansion of Eqs. (\ref{eqXY}) around the steady state, with $\wef = \we + \delta \omega (t)$, $X = \overline{X}+\delta X$ and $Y = Y_0 + \delta Y$:  
\begin{subequations}
\label{eqXYa}
\begin{align}
\delta \dot{X}+\frac{\gef}{2} \delta X - \delta \omega(t) \,Y_0 = \frac{1}{M \we}\left[f^{(1)}_T + f^{(1)}_s + f^{(1)}_{BA} \right]    \\
\delta \dot{Y}+\frac{\gef}{2} \delta Y = \frac{1}{M \we}\left[f^{(2)}_T + f^{(2)}_s + f^{(2)}_{BA} \right]    \\
\delta\omega(t) = \delta \woef(t) - \int_{-\infty}^{t} \,\mathcal{\bar{G}}(t,t')\, \left[\,\overline{X}(\delta\omega(t'))+\delta X(t')+n^{(1)}(t')\right] \,\mathrm{d}t'   \, .
\end{align}
\end{subequations}
We have few important remarks on the above relations. The first one is that the equation governing the fluctuations of the $Y$ quadrature is the same that we would have without feedback, therefore $\delta Y$ behaves as in a free oscillator and, in particular, it can be used to reliably measure the external force. Second point, we have a well defined phase plane: the oscillator is not just frequency stabilized, but also phase-locked to the reference. Third issue, the response function of the $Y$ quadrature is stable, with a peak frequency defined \emph{a priori} (at $\omega=0$, corresponding to $\we$ for the evolution of $x$) and, as a consequence, stable width $\gef$ and peak signal-to noise ratio. Such parameters stability is very important for an easier application of optimal filtering.  

The spectrum of the measured $Y_m$ quadrature calculated from Eq. (\ref{eqXYa}b) can be written in the form
\begin{equation}
S_{Ym} = \lor(\omega)\, S_F + S_n/2
\label{eqSY}
\end{equation}
with
\begin{equation}
\lor(\omega)=\mathcal{A} \frac{\gef}{\omega^2 + \left(\frac{\gef}{2}\right)^2} \, .
\label{eqlor}
\end{equation} 
where $\mathcal{A}=\int_{-\infty}^{\infty}\lor(\omega)\,\mathrm{d}\omega/2\pi = 1/(2 \gef \,M^2 \,\we^2)$
and the total force noise spectral density is $S_F = S_s+S_T+S_{BA}$.

The treatment of this Section includes slow fluctuations of $\woef$ as well as its fast, although weak variations that can be considered as phase fluctuations. The case of strong and fast variations of $\woef$, producing trajectories in the phase plane that take $\mathbf{R}$ far from the region with $\theta < 1$, requires a numerical integration of Eqs. (\ref{eqXY}) and the approximation of a free $Y$ quadrature is no more reliable.

By excluding the coherent excitation and the frequency control, the spectrum of both quadratures, for an opto-mechanical resonance at $\woef = \we+\delta\omega$, is 
\begin{equation}
S_{Xm} = S_{Ym} = \frac{1}{2}[\lor(\omega-\delta\omega)+\lor(\omega+\delta\omega)] S_F +\frac{S_n}{2}
\label{duelor}
\end{equation}
and, in case of slow fluctuations of $\delta\omega$, the spectral peaks assume the shape of a Voigt profile, maintaining a constant area.

\subsection{Force measurement strategies}

We consider two possible measurement strategies, with the aim of detecting a weak stochastic signal force $f_s$ hidden by the thermal background. In other  words, we are seeking for a precise measurement of the stochastic force in order to resolve its weak variations due to changes in $S_s$. We are not dealing with measurement accuracy and reproducibility, that both depend critically on absolute calibrations. 

The first strategy is simply measuring the area $\sigma^2$ of the resonance peak. The advantage of this method is that frequency stability of the opto-mechanical oscillator is not crucial: the peak area can be calculated by direct integration of the spectrum of $x$ within an appropriate frequency interval, provided that $\wef$ is well within the integration band, and the latter is extended to few $\gef$ yet maintaining a negligible contribution of the background noise $S_n$. The same measurement can be performed, with equal efficiency, on the spectrum of a quadrature. The estimated force spectral density is $E\{S_F\} = \sigma^2/\mathcal{A}$. The drawback of this method is the rather slow improvement of the statistical uncertainty, decreasing as $\propto \sqrt{\tau_c/t_{meas}}$ where the correlation time is now $\tau_c=1/\gef$. The reason is that this strategy does not exploit the full information contained in the signal, whose spectrum around resonance is dominated by the effect of the force fluctuations even well beyond the width $\gef$.  

\begin{figure} [h]
\centering
\includegraphics[width=0.9\textwidth]{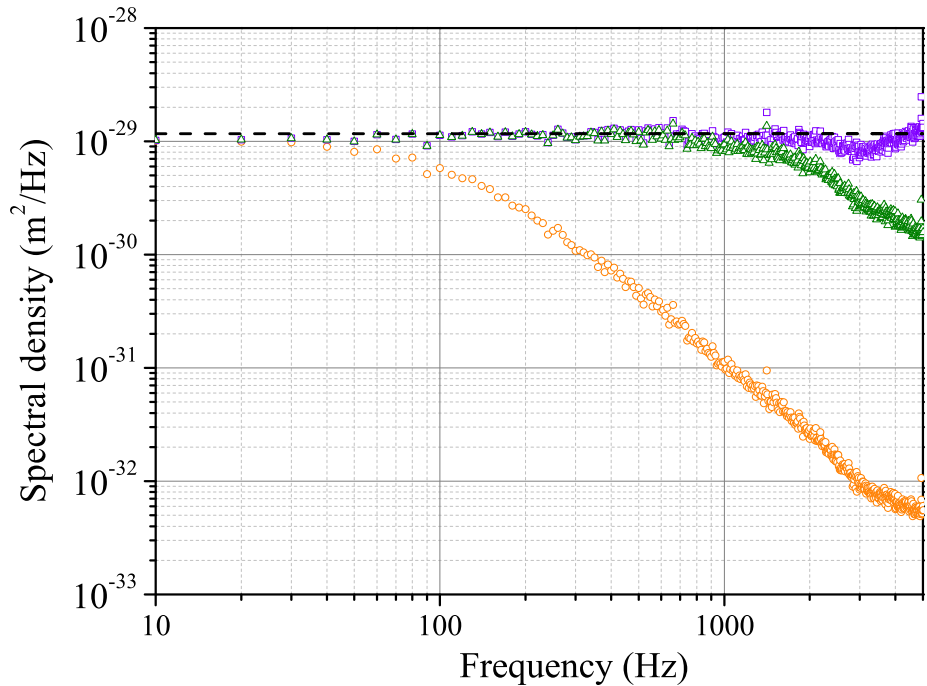}
\caption{(Color online) Measured spectral density in the $Y$ quadrature ($S_{Ym}$) (orange circles); whitened spectrum (violet squares); with complete Wiener filtering (green triangles).}
 \label{figWiener}
\end{figure}

The second strategy is a close approximation of the Wiener filtering, that represents the optimal choice in case of stationary noise. The non-causal Wiener filter, applied to the spectrum $S_{Ym}$ of Eq. (\ref{eqSY}), is defined as 
\begin{equation}
|W(\omega)|^2 = \frac{1}{\lor(\omega)}\left[\frac{1}{1+\Gamma\frac{\lor(0)}{\lor(\omega)}}\right]^2
\label{eqwiener}
\end{equation}
and the maximum information on $S_F$ from the experimental $S_{Ym}$ is obtained from the filtered spectrum $S_W = |W|^2 S_{Ym}$. The $1/\lor$ factor in Eq. (\ref{eqwiener}) is a whitening and calibration function, while the term between square brackets is a weight function that requires preliminary estimate of the noise-to-peak-signal ratio $\Gamma$. Its optimal value is $\Gamma_{opt} = S_n/2 \lor(0) S_F$, but an efficient, even if sub-optimum, filter can choose a $\Gamma > \Gamma_{opt}$ \cite{Astone1990}. In any case, preliminary fit of a spectrum $S_{Ym}$ allows to extract the parameters $\gef$ and $\Gamma$ for the following application of the Wiener filtering procedure. The correlation time of the filtered signal is now $\tau_c \sim \sqrt{\Gamma}/\gef$, yielding  a faster improvement of the statistics with $t_{meas}$ with respect to the previous strategy. For an optimum filter (with $\Gamma = \Gamma_{opt}$), $1/\tau_c$ corresponds to the effective sensitivity bandwidth, i.e., to the frequency band where the effect of force noise falls below the measurement sensitivity (i.e., $\lor(\omega) S_F = S_n/2$). An example of the application of the whitening function and the complete Wiener filter to a real spectrum is shown in Fig. \ref{figWiener}. The force spectral density is estimated by integrating the filtered spectrum $S_W$ and dividing the result by the effective bandwidth $\sim 1/\tau_c$. In our real data some spurious peaks appear in the spectrum at few kHz from the opto-mechanical resonance, therefore the integration is truncated at $\omega_{cut}/2\pi$=3kHz, slightly below $1/\tau_c$. More details on the choice of $\omega_{cut}$ and on the consequent effective bandwidth are reported in the Appendix.

As we have seen, the application of the Wiener filtering requires the knowledge of the transfer function between force noise and output. For this reason, the parametric control strongly facilitates the filtering procedure, by fixing both the opto-mechanical resonance frequency at $\wef = \we$ and, as a consequence, its width $\gef$. Without control, optimal filtering would require an adaptive tuning of the parameters, that we are not trying to apply in this work. 

\section{Experiment}

\subsection{Experimental apparatus}

A sketch of our experimental system is shown in Fig. \ref{schema}c. A Fabry-Perot cavity is formed between a micro-oscillator with high reflectivity dielectric coating as end mirror and a standard concave input coupler. The cavity length is 0.57~mm and its finesse is 57000 (half-linewidth $\kappa/2\pi = 2.3$~MHz). The input coupler is glued on a piezo-electric transducer for coarse tuning, and the cavity is kept in a vacuum chamber at $10^{-3}$~Pa. The low-deformation micro-mirror \cite{SerraAPL2012,SerraJMM2013} has resonance frequency $\wm/2\pi =128960$~Hz, mechanical quality factor $Q=\wm/\gm=16000$ (limited, at room temperature, by thermoelastic losses) and effective mass $M = 1.35~10^{-7}$~Kg. More details on the measurements of the opto-mechanical parameters are reported in Refs. \cite{SerraAPL2012,SerraPRA2012}. 

Two laser beams derived from the same Nd:YAG source are overlapped with orthogonal polarizations and optically matched to a cavity longitudinal mode with an efficiency of $\approx96\%$. From the reflected first beam (\emph{signal beam}, with a power of $80 \mu$W) we obtain a dispersive profile of the optical resonance (PDH signal) through phase modulation at 13.3~MHz and phase-sensitive detection \cite{Drever}. Such signal is exploited for locking the laser beam to the cavity resonance. Moreover, in the approximately linear region around resonance, the PDH signal is proportional to the oscillator displacement and is used both for monitoring its motion and in the parametric control loop described below. We remark that the bandwidth of the laser locking is kept at $\sim30$~kHz (well below the mechanical frequency) and additional strong notch filters assure that the laser frequency servo loop has no effect in the frequency region of interest (around the mechanical resonance).

The second beam (\emph{control beam}), with a power of 1 mW at the cavity input, is frequency shifted with respect to the signal beam, and is used to control the optical spring. The adjustable frequency shift compensates the cavity birefringence and determines the detuning of the control beam with respect to the cavity resonance. The ratio between opto-mechanical frequency shift due to the optical spring and control beam detuning is $8 \cdot 10^{-3}$ Hz/Hz. In addition, an electro-optic intensity modulator imposes a weak sinusoidal modulation in the power of the control beam and consequently in the radiation pressure acting on the micro-mirror. 
 
The PDH signal is calibrated by means of a modulation at $\sim20$kHz sent to the laser frequency controller. The amplitude of this modulation at the input of the controller is directly measured during the acquisitions (since it is influenced by the frequency servo loop). This measurement, as well as the measurement of the amplitude of the corresponding modulation in the PDH signal, are repeated every 1s during the data acquisition, in order to compensate for (weak) changes in the detection efficiency. The laser tuning rate (in Hz/V) had been previously calibrated with a Michelson interferometer, and the ratio between the laser frequency and the cavity length allows to convert the detuning into cavity displacement. The overall calibration has an absolute accuracy of $\sim20\%$ (we point out that such uncertainty in the calibration factor do not influence the possibility to resolve weak signal variations, that is the object of this work). 

The PDH signal is also sent to a double-phase, digital lock-in amplifier and integrated with a time constant of 80$\mu$s. For the configuration with parametric control of the opto-mechanical frequency, the lock-in oscillator is sent to the intensity modulator of the control beam. A preliminary scan of its frequency allows to reconstruct the response function of the mechanical oscillator and to tune the phase of the lock-in amplifier in order to have the dispersive component at the $X$ output. The reference oscillator is then set to 127400 Hz and the $X$ output of the lock-in amplifier is integrated and sent to the drivers of the acousto-optic modulators that vary the detuning of the control beam. The opto-mechanical resonance is now phase-locked to the reference oscillator. The detuning of the control beam corresponds to about $0.09 \kappa$ and the oscillator is in rather strong optical damping condition, with a resonance width of $\gef/2\pi \simeq 200$Hz. For the configuration without parametric control, the effective opto-mechanical frequency is moved to about 127400 Hz by hand tuning the control beam, but the lock-in reference frequency is set at 127200 Hz, so that the well defined resonance peak at $\sim200$ Hz allows to measure more accurately its parameters.

\subsection{Measurements}

The signal from the $Y$ output of the lock-in amplifier is acquired by a digital scope with a resolution of 12 bit and a sampling interval of 21$\mu$s. Data are acquired by the scope in 35 consecutive time traces, each one lasting about 20 seconds (corresponding to $\sim 10^6$ data points) covering in all nearly 12 minutes, then stored in a hard disk. Several of such series are taken separated by periods of few minutes (necessary to write the data on disk), for a total observation time of several tens of minutes. 

\begin{figure} [h]
\centering
\includegraphics[width=0.9\textwidth]{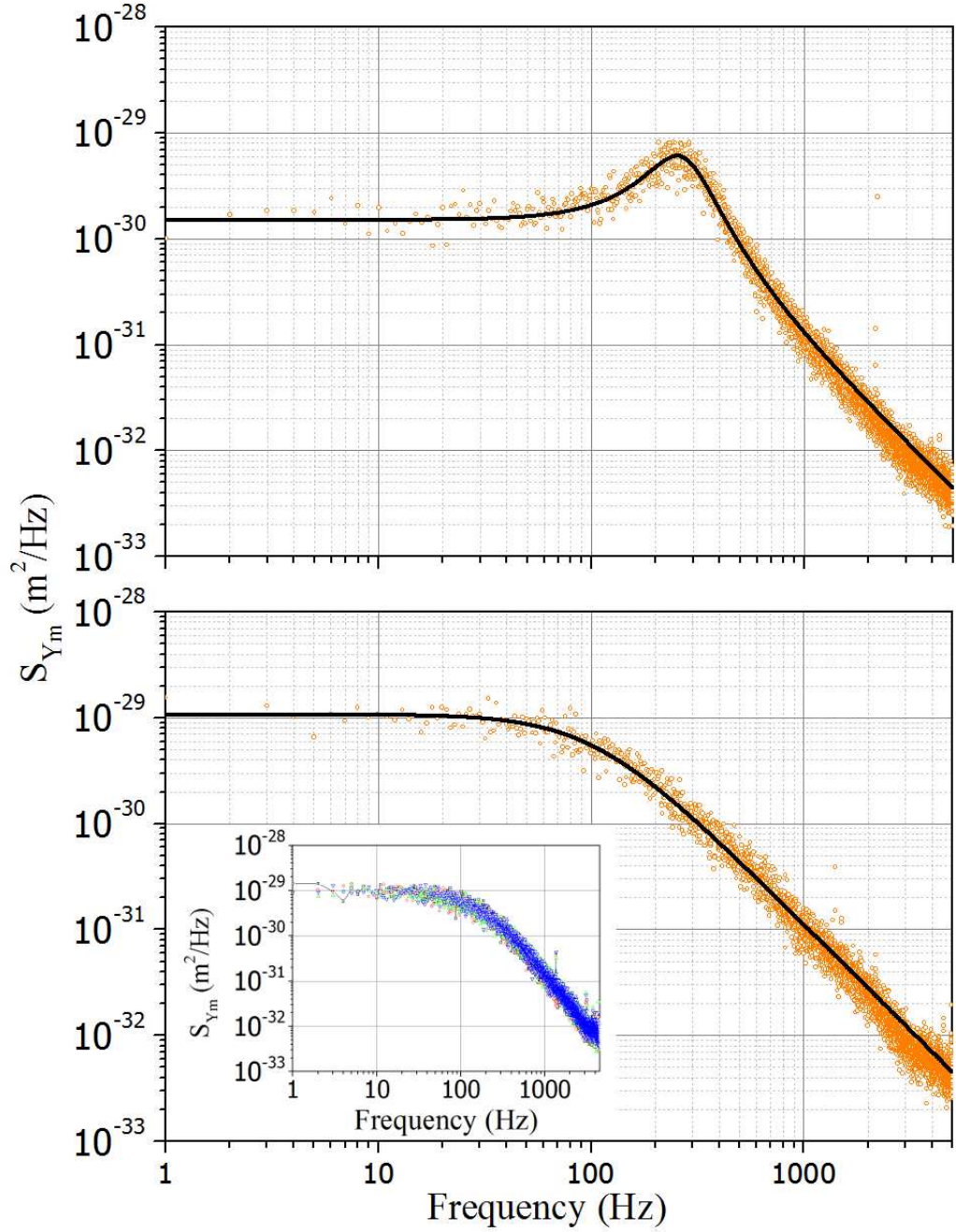}
\caption{(Color online) Spectral densities of the $Y$ quadrature ($S_{Ym}$), for an oscillator without (upper panel) and with (lower panel) parametric control. With a solid line we show the respective fitting functions. In the inset, we compare spectra obtained with different values of the parametric control gain, showing that the control do not influence the dynamics of the $Y$ quadrature.}
 \label{spettri}
\end{figure}

The time series are divided into 100ms long segments, a duration much larger than their correlation time. For each section the power spectrum is calculated using a FFT algorithm, and corrected for the transfer function of the lock-in amplifier. The spectra corresponding to the first 20 seconds are averaged, and the averaged spectrum is fitted to Eq. (\ref{eqSY}) (when the parametric control is active) or to Eq. (\ref{duelor}) (without control). An example of the averaged spectra and the fits are shown in Fig. \ref{spettri}. From the fitting procedure we obtain the resonance width, the signal maximum and, in the absence of the control, also the resonance frequency. The signal maximum $Max$ is just exploited to define the value of the parameter $\Gamma$ to be used for Wiener filtering. At this purpose, we consider a conservative value of the background additive noise on $Y$, at $S_{BG}=8\cdot 10^{-33} $m$^2$/Hz (one order of magnitude larger than the real $S_n$) and define $\Gamma=S_{BG}/Max$. A typical value of $\Gamma$ is $10^{-3}$.

\begin{figure} [h]
\centering
\includegraphics[width=0.9\textwidth]{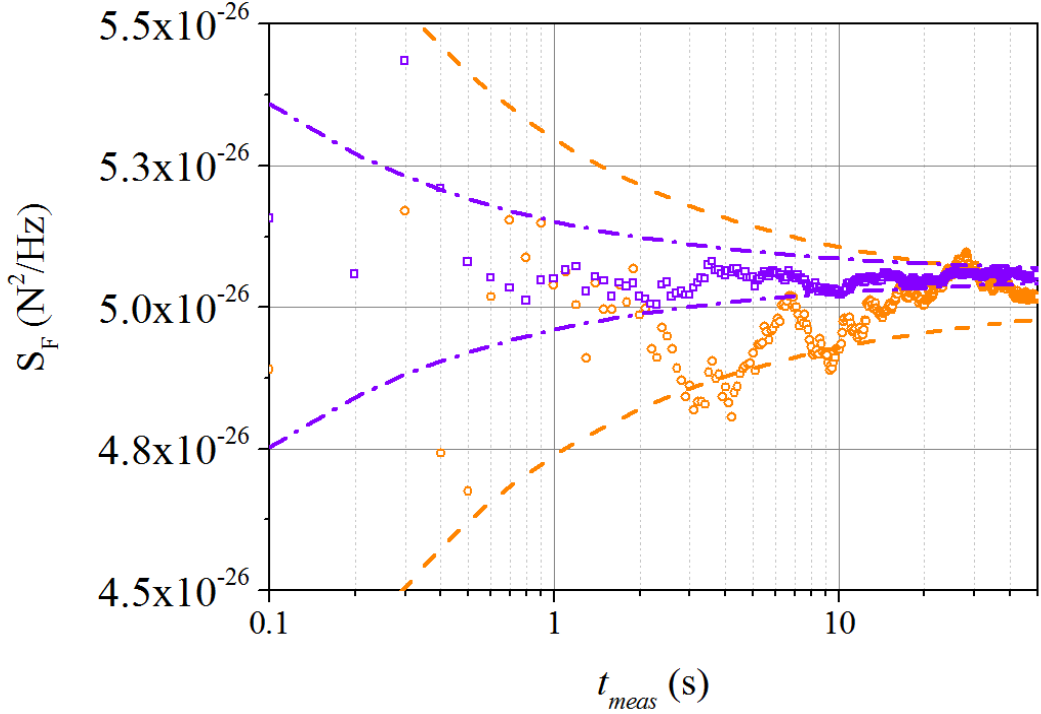}
\caption{(Color online) Average over a measurement time $t_{meas}$ of the force noise spectral density $S_F$, measured on the oscillator with parametric control using the peak area (orange circles) and the Wiener filtered   spectra (violet squares). The confidence bands (respectively dashed and dash-dotted lines), corresponding to one standard error, are calculated in the Appendix.}
 \label{media}
\end{figure}

From each of the following spectra (after the first 20s) we calculate the force spectral density $S_F$ using the different methods described in the previous Section (i.e., from the peak area and using Wiener filtering, both in the configuration with parametric feedback and with free-running oscillator). We report in Fig. \ref{media} the average $\bar{S}_F(t_{meas})$ of $S_F$ accumulated over $m$ consecutive spectra, corresponding to a measurement time $t_{meas} = m \tau$, where $\tau = 100$ms is the time interval used for calculating each spectrum. The relative standard error is given by $\sigma_{\mathrm{REL}} \simeq 2/\sqrt{t_{meas} \gef}$ for the measurement with the peak area, and $\sigma_{\mathrm{REL}} \simeq \sqrt{2\pi/t_{meas}\omega_{cut}}$ when using Wiener filtered data (these expressions refer to the configuration with parametric control where the peak is centered at null frequency, and the latter relation is valid for $\omega_{cut} \ll \gef/2\sqrt{\Gamma}$; exact calculations are reported in the Appendix). $\sigma_{\mathrm{REL}}$ is used to calculate the confidence regions $(1 \pm \sigma_{\mathrm{REL}})\bar{S}_F$, where $\bar{S}_F$ is the average at the end of the measurement period. The figure shows the expected convergence of the measured $\bar{S}_F(t_{meas})$, which is clearly faster for the filtered data.  

\begin{figure} [h]
\centering
\includegraphics[width=0.9\textwidth]{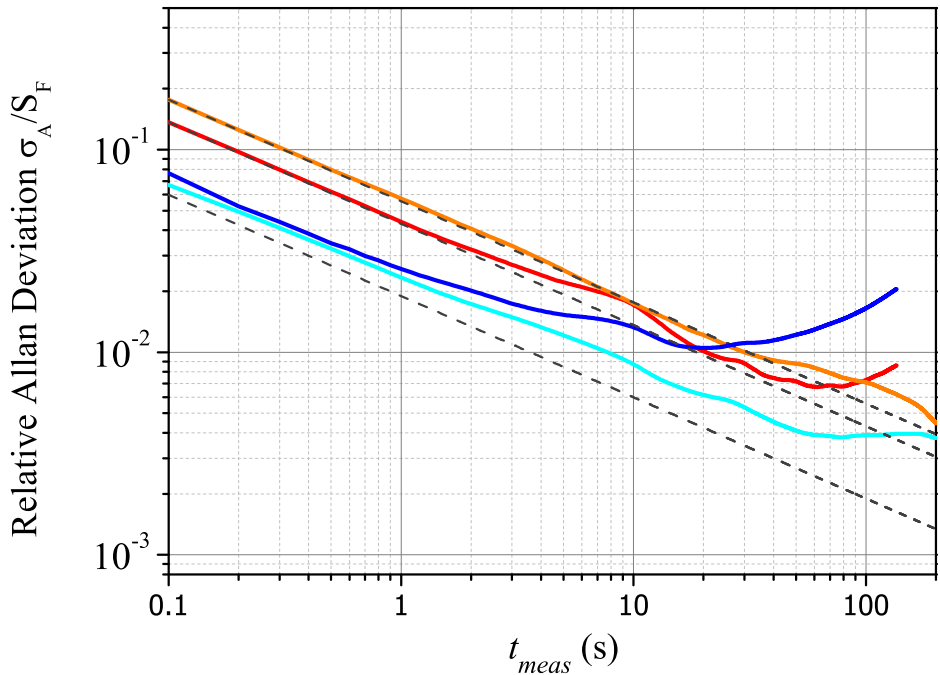}
\caption{(Color online) Relative Allan deviation concerning the measurement of the input stochastic force $S_F$, performed with four different procedures. Solid lines, from the upper to the lower curve (as seen in the left region of the graph): measurement from the peak area, with parametric control (orange); the same, without control (red); measurement from the Wiener-filtered data, without parametric control (deep blue); the same, with control (light blue). Dashed lines display the expected behaviour in the absence of long-term effects, given by Eq. (\ref{A3}) (upper line), Eq. (\ref{A3bis}) (middle line), and Eq. (\ref{A5}) (considering an implementation of the optimal filter; lower line).}
 \label{figAllan}
\end{figure}

The calculation of the confidence region reported in Fig. \ref{media} is just valid for a stationary system. A more reliable assessment on the measurement stability on the long term and on the achievable resolution is provided by the Allan variance \cite{Allan}. In our case, its estimator is defined as 
\begin{eqnarray}
\sigma_A^2(m) = \frac{1}{N-m}\sum_{k=1}^{N-2m+1} \frac{\left(\bar{x}_{k+m}-\bar{x}_k\right)^2}{2} \\
\bar{x}_k(m) =\frac{1}{m} \sum_{n = k}^{k+m-1} S_F (n)
\end{eqnarray}   
where $S_F (n)$ is the value of force spectral density calculated from the $n$th spectrum and $N$ is the total number of spectra. The Allan deviation $\sigma_A(m)$ estimates the one sigma uncertainty that can be obtained with a measurement lasting $t_{meas} = m \tau$. The calculated relative Allan deviation (i.e., $\sigma_A$ divided by $\bar{S}_F$) is reported in Fig. \ref{figAllan} for the different measurement strategies. We can derive two main considerations: a) the measurement with Wiener filtering improves the statistical uncertainty much faster than the measurement from the peak area. For the former, a $1\%$ resolution is obtained after 10s and the best resolution of $0.4\%$ is achieved, thanks to the parametric stabilization, after one minute; for the latter, the necessary measurement periods are about three times longer, in agreement with the ratio between the respective $\sigma_{\mathrm{REL}}$; b) for measurement periods exceeding 1s, the parametric control is crucial for the application of Wiener filtering. The measurement resolution does not improves any more after one minute: with the parametric control it remains constant, while it becomes even worse without control. It means that the parametric control also allows a much more relaxed choice of the optimal measurement time.    

\section{Conclusions}

We have analyzed different possible procedures for measuring the stochastic force acting on a micro opto-mechanical system. In particular, we have compared the usual strategy based on the direct measurement of the area of the resonance peak (or, equivalently, of the variance in the oscillator displacement) with a more refined data analysis that approaches the optimal Wiener filtering. For the latter case, we have introduced an abrupt bandwidth limitation that allows a near-optimal realistic measurement procedure. We have shown that, while for the former method the optical damping, decreasing the oscillator coherence time, can improve the resolution of the measurement in a given observation time, the appropriate filtering gives sensibly better results which are mostly independent on such coherence time.

The implementation of the Wiener filtering is greatly facilitated and more effective by using a parametric control of the oscillator frequency, a technique that we have recently introduced and that we have analyzed here in details. Thanks to such active stabilization, our system can reliably detect variations of the stochastic force below
1$\%$ within one minute. We remark that a correct assessment of the really achievable resolution with long integration periods cannot be simply based on the convergence of the averaged measurement. Indeed, such indicator underestimates the effect of system long-term instabilities and parameter drifts. Using the Allan variance as correct estimator, we show that parametric control plays a crucial role in the achieved performance.

The procedure for the measurement of the stochastic force that we have described in this work, including optimal filtering and parametric control, can be applied in a large variety of micro- and nano-mechanical systems, including those based on electric measurements and microwave radiation. Detecting a weak stochastic signal on a stronger background is an important task in the research field of quantum mechanics with macroscopic oscillators, in particular when exploring the properties of oscillators with low occupation number, or, e.g., in a squeezed state \cite{Pontin2013,Clerk2008,Hertzberg2009,Bowen2013} or other peculiarly quantum states. In this situation, the measurement back-action can destroy the interesting features. Particular measurement schemes can be conceived and applied \cite{chan11,Clerk2008,Hertzberg2009,Kronwald2013}, but the use of a weak measurement, where the signature of the oscillator is intrinsically weaker than the measurement noise (see, e.g., in Ref. \cite{Thompson08}), can be a useful affordable solution. The procedures investigated in this work would thus provide a valuable help.

\section{Acknowledgments}

F.M. thanks M. Prevedelli for the discussion on phase locking. This work has been supported by the European Commission (ITN-Marie Curie project cQOM), by MIUR (PRIN 2010-2011) and by INFN (HUMOR project).

\newpage

\appendix
\section{}

\begin{figure} [h]
\centering
\includegraphics[width=0.9\textwidth]{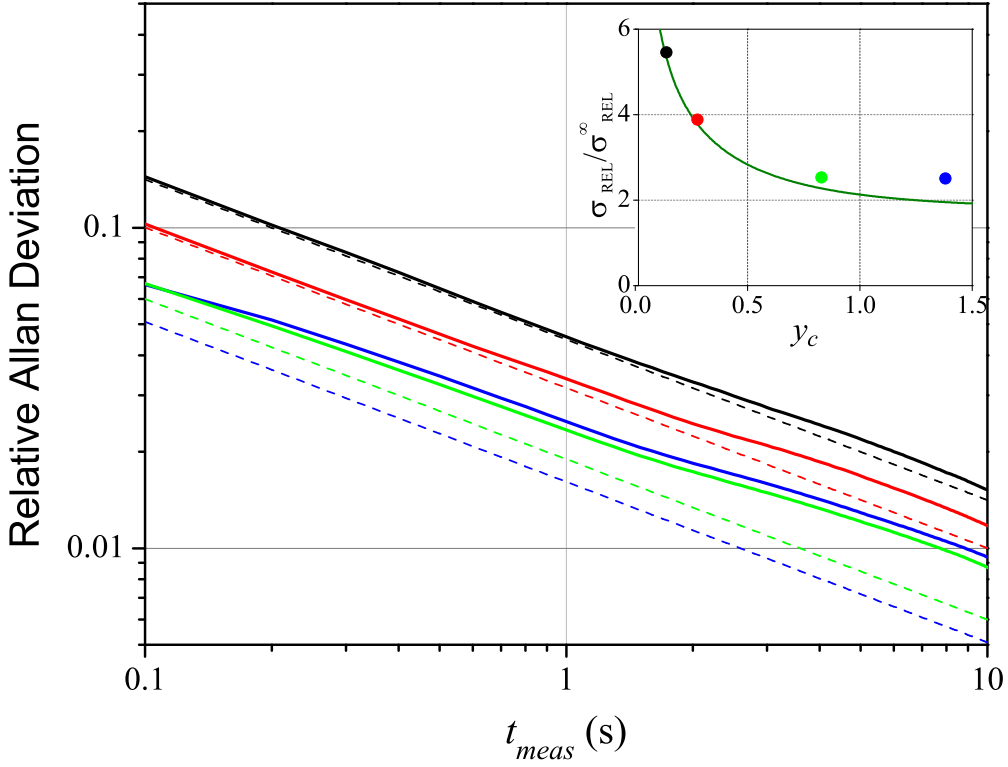}
\caption{(Color online) Relative Allan deviation for the measurement of $S_F$ using the Wiener filtered data, for different values of the cutoff frequency. The dashed lines correspond to the calculated $\sigma_{\mathrm{REL}}$ for $\omega_{cut}/2\pi = $500 Hz, 1 kHz, 3 kHz and 5 kHz (from the upper to the lower line). The solid curves are the experimental results, and we see that the data extracted with the cutoff at 5 kHz overtake the curve corresponding to  $\omega_{cut}/2\pi = $3 kHz. In the inset, the experimental relative Allan deviation at $t_{meas} = 0.1$s,  normalized to the corresponding $\sigma_{\mathrm{REL}}^{\infty}$, is reported for the same values of $\omega_{cut}$ and compared with the theoretical behavior (calculated form Eqs. (\ref{A5}-\ref{A6})) shown with a solid line.}
\label{figAppendix}
\end{figure}

We consider a Gaussian, zero mean stochastic process $x(t)$ with finite variance $\sigma_x^2$, correlation function $C_{xx}$, and power spectral density $S_{xx}$. The estimate of the mean square of $x(t)$ in the interval $[0,t_{meas}]$ has expectation value $\sigma^2_x$ and standard deviation \cite{Astone1990,Bendat2010}
\begin{equation}
STD \simeq \left[\frac{2}{t_{meas}}\int_{-\infty}^{\infty} C_{xx}^2(\tau) \mathrm{d}\tau\right]^{\frac{1}{2}} \,.
\label{A1}
\end{equation}
The relative standard deviation is defined as $\sigma_{\mathrm{REL}}=STD/\sigma_x^2$, and it can be expressed in terms of spectral densities using
\begin{subequations}
\label{A2} 
\begin{align}
\sigma_x^2 = \int_{-\infty}^{\infty} S_{xx} (\omega) \frac{\mathrm{d}\omega}{2\pi}   \\
STD \simeq  \left[\frac{2}{t_{meas}}\int_{-\infty}^{\infty} 
S_{xx}^2(\omega) \frac{\mathrm{d}\omega}{2\pi}\right]^{\frac{1}{2}} \,.
\end{align}
\end{subequations}
For a spectrum given by $S_{xx}(\omega) \propto \lor(\omega)$ we obtain the relative standard deviation \cite{Astone1990}
\begin{equation}
\sigma_{\mathrm{REL}}=\frac{2}{\sqrt{t_{meas} \gef }}  \, .
\label{A3}
\end{equation}
This expression can be used for the relative uncertainty in the measurement of $S_F$ using the peak area, since in this case we can neglect the measurement noise $S_n$ and the finite integration band defined by $\omega_{cut}$. For the Wiener-filtered process, using Eqs. (\ref{eqSY}), (\ref{eqwiener}) and the expression of $S_W$ and $\Gamma_{opt}$  we can write the output spectrum in the form
\begin{equation}
S_{xx} \propto \lor(\omega) \frac{\lor(\omega)+\lor(0)\Gamma_{opt}}{\left(\lor(\omega)+\lor(0)\Gamma\right)^2} \,.
\end{equation}
and the relative standard deviation as 
\begin{subequations}
\label{A5} 
\begin{align}
\sigma_{\mathrm{REL}}=\frac{2}{\sqrt{t_{meas} \gef }}\left(\frac{\Gamma}{1+\Gamma}\right)^{\frac{1}{4}}\frac{\sqrt{\pi\int_0^{y_c}\left[\frac{1+g y^2}{\left(1+y^2\right)^2}\right]^2\,\mathrm{d}y}}{\int_0^{y_c}\frac{1+g y^2}{\left(1+y^2\right)^2}\mathrm{d}y}  \\
y_c = \omega_{cut} \frac{2}{\gef}\sqrt{\frac{\Gamma}{1+\Gamma}}  \\
g = \frac{\Gamma_{opt}(1+\Gamma)}{\Gamma(1+\Gamma_{opt})}  \, .
\end{align}
\end{subequations}
It is useful to consider the two limits $y_c \to \infty$ and $y_c \ll 1$, that for  $g \ll\ 1$ and $\Gamma \ll 1$ can be written respectively as 
\begin{equation}
\sigma_{\mathrm{REL}}^{\infty} \simeq \sqrt{\frac{10}{t_{meas} \gef/\sqrt{\Gamma}}}
\label{A6}
\end{equation}
and 
\begin{equation}
\sigma_{\mathrm{REL}} \simeq \sqrt{\frac{2\pi}{t_{meas} \omega_{cut}}}  \, .
\end{equation}
In the inset of Fig. \ref{figAppendix} (solid line) we show the behavior of $\sigma_{\mathrm{REL}}/\sigma_{\mathrm{REL}}^{\infty}$ as a function of $y_c$. The relative accuracy is just $20\%$ worse if the integration is limited to $y_c=1$ and, on the other hand, a frequency cutoff allows to reject spurious signals that can appear around the interesting resonance.

For a spectrum formed by a couple of symmetric Lorentzian peaks centered at $\pm \delta\omega$ (see Eq. (\ref{duelor})), the relative standard deviation, when measuring directly the peaks area, becomes
\begin{equation}
\sigma_{\mathrm{REL}}=\frac{2}{\sqrt{t_{meas} \gef }}\,\sqrt{\frac{\gef^2+2 \delta\omega^2}{\gef^2+4 \delta\omega^2}}  \, .
\label{A3bis}
\end{equation} 
The Wiener filter is obtained from the expression for a single peak, given in Eq. (\ref{eqwiener}), by replacing $\lor(\omega) \to 0.5 (\lor(\omega-\delta\omega)+\lor(\omega+\delta\omega))$. Due to the spectral flattening action of the Wiener filter, the filtered output is very similar to the case of the single peak. As a consequence, for our typical parameters, the two theoretical values of $\sigma_{\mathrm{REL}}$ differ by less than $1\%$.

The relative Allan deviation is equal to $\sigma_{\mathrm{REL}}$ in the absence of excess fluctuations (typically, for short measurement time). In our experiment, $y_c = 1$ for $\omega_{cut}/2\pi \simeq 6300$ Hz. In Fig. \ref{figAppendix} we report the measured relative Allan deviation as a function of $t_{meas}$ for different values of the cutoff frequency, together with its expected behavior. When $\omega_{cut}/2\pi$ surpasses 3 kHz, the presence of additional peaks starts to influence the measurement. This is also visible in the inset, where the measured relative Allan deviation at $t_{meas} = 0.1$ s, normalized to the calculated $\sigma_{\mathrm{REL}}^{\infty}$, is reported for different values of $\omega_{cut}$ and compared with the theoretical behavior.

\end{document}